# An Adaptive Technique Using Advanced Encryption Standard To Implement Hard Disk Security

Minal Moharir, Dr. A V Suresh,
R V College of Engg., Bangalore-59,India

*Abstract:* The main objective of the paper is to study and develop an efficient method for Hard Disk Drive(HDD) Security using Full Disk Encryption (FDE) with Advanced Encryption Standards(AES) for data security specifically for Personal Computers(PCS) and Laptops.

The focus of this work is to authenticate and protect the content of HDD from illegal use. The paper proposes an adaptive methods for protecting a HDD based on FDE. The proposed method is labeled as DiskTrust. FDE encrypts entire content or a single volume on your disk.

DiskTrust implements Symmetric key cryptography with, Advanced Encryption Standards.

Finally, the applicability of these methodologies for HDD security will be evaluated on a set of data files with different key sizes.

*Keywords: Information Security, Integrity, confidentiality, Authentication, Encryption.*

## 1 INTRODUCTION

As of January 2011 the internet connected an estimated 941.7 million computers in more than 450 countries on every continent, even Antarctica (Source: Internet Software Consortium's Internet Domain Survey; www.isc.org/index.pl). The internet is not a single network, but a worldwide collection of loosely connected networks that are accessible by individual computer hosts, in a variety of ways, to anyone with a computer and a network connection. Thus, individuals and organizations can reach any point on the internet without regard to national or geographic boundaries or time of day.

However, along with the convenience and easy access to information come risks. Among them are the risks that valuable information will be lost, stolen, changed, or misused. If information is recorded electronically and is available on networked computers, it is more vulnerable than if the same information is printed on paper and locked in a file cabinet. Intruders do not need to enter an office or home; they may not even be in the same country. They can steal or tamper with information without touching a piece of paper or a photocopier. In this way security of stored information is an important issue. The proposed paper consider the security of Hard Disk Drive which is a fundamental element in computing chain.

The paper organized as follows. Related work, gap & problem is described in Section 2. A view of simulation and experimental design is given in section 3. Simulation results are shown in section 4. Finally the conclusions are drawn section 5.

## 2 RLATED WORK

The related survey is divided into two parts. The first part is survey about full disk encryption. The second part is survey about advanced encryption standards.

Information security is the process of protecting information. It protects its availability, privacy and integrity. More companies store business and individual information on computer than ever before. Much of the information stored is highly confidential and not for public viewing. Without this information, it would often be very hard for a business to operate. Information security systems need to be implemented to protect this information. There are various ways to implement Information security systems. One of the popular technique is full disk encryption. **F**ull **D**isk **E**ncryption (FDE) is the safest way to protect digital assets, the hard drive is a critical element in the computing chain because it is where sensitive data is stored. Full disk encryption increases the security of information stored on a laptop significantly. It helps to keep business critical data absolutely confidential. Moreover, full disk encryption helps to meet several legislative requirements. Various techniques to implement FDE are discussed as follows: **Tabel1**:

| Name | Developed | Released | Licence | OS |
|---|---|---|---|---|
| TrueCrypt | TrueCrypt Foundation | 2009 | Free | Linux, Windows |
| Discryptor | Cosect | 2008 | Commercial, closed source | Windows, Vista |
| DriveSentry | DriveSentry | 2008 | Commercial, closed source | Windows, Vista |
| R-Crypto | R-Tools Technology Inc | 2008 | Free, closed source | Windows XP, Vista |

The second part of survey covers implementation of Encryption Algorithms. Many encryption algorithms are widely available and used in information security. They can be categorized into Symmetric (private) and Asymmetric (public) keys encryption. In Symmetric keys encryption or secret key encryption, only one key is used to encrypt and decrypt data. The key should be distributed before transmission between entities. Keys play an important role. If weak key is used in algorithm then every one may decrypt the data. Strength of Symmetric key encryption depends on the size of key used. For the same algorithm, encryption using longer key is harder to break than the one done using smaller key. Brief definitions of the most common encryption techniques are given as follows:  DES: (Data Encryption Standard), was the firstencryption standard to be recommended by NIST (National Institute of Standards and Technology).DES is (64 bits key size with 64 bits block size) . Since that time, many attacks and methods recorded the weaknesses of DES, which made it an insecure block cipher [3],[4].3DES is an enhancement of DES; it is 64 bit block size with 192 bits key size. In this standard the encryption method is similar to the one in the original DES but applied 3 times to increase the encryption level and the average safe time. It is a known fact that 3DES is slower than other block cipher methods [3]. RC2 is a block cipher with a 64-bits block cipher with a variable key size that range from 8 to128 bits. RC2 is vulnerable to a related-key attack using 234 chosen plaintexts [3]. Blowfish is block cipher 64-bit block - can be used as a replacement for the DES algorithm. It takes a variablelength key, ranging from 32 bits to 448 bits; default 128 bits. Blowfish is unpatented, license-free, and is available free for all uses. Blowfish has variants of 14 rounds or less. Blowfish is successor to Twofish [5]. AES is a block cipher .It has variable key length of 128, 192, or 256 bits; default 256. it encrypts data blocks of 128 bits in 10, 12 and 14 round depending on the key size. AES encryption is fast and flexible; it can be implemented on various platforms especially in small devices[6]. Also, AES has been carefully tested for many security applications [3], [7]. RC6 is block cipher derived from RC5. It was designed to meet the requirements of the Advanced Encryption Standard competition. RC6 proper has a block size of 128 bits and supports key sizes of 128, 192 and 256 bits. Some references consider RC6 as Advanced Encryption Standard [8].

*2.1 Research Gap*

The FDE technology discussed in above survey are encrypting the entire contents of Hard disk Drive. However encryption of the entire HDD is expensive in terms of time and cost.  DiskTrust, the technology proposed here, creates a hidden volume on HDD, which is not visible, accessible to the unauthorized user. The data store in this hidden volume is encrypted using robust(Rijndael) encryption algorithm.  In this way DiskTrust technology follows CIA properties of secure information along with hidden partition.

2.2 Problem Definition

DiskTrust technology implements security on the hard drive itself, to provide a foundation for trusted computing.

The technical objectives of the paper are:

1. Create Hidden partition
2. Execute issuance protocol to check authentication.
3. Execute encryption/decryption algorithm while reading /writing data on Hard Disk Drive.

3 SIMULATION AND DESIGN:

This section describes some of the important results that were found as part of the implementatton.

*3.1 Implementation of Hidden Volume*

DiskTrust Security user interfaces are shown below in the screenshots. The user interface is basically a frame work application where user can use the application

*3.1.1 Main Application Window*

The Main application window holds multiple options such as CreateVolume, Mount and Dismount All.

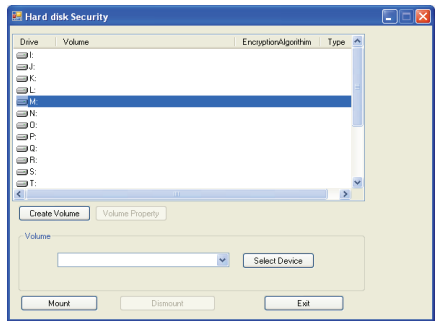

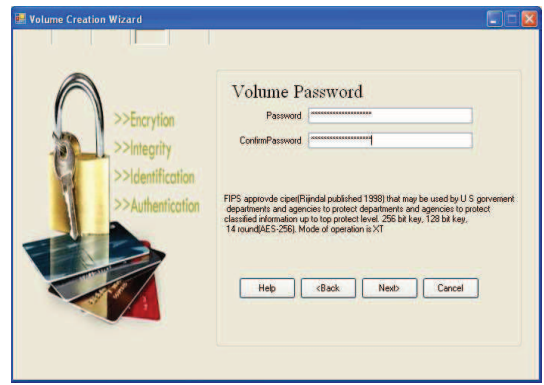

Figure 3.2.1 Screenshot **of Volume Password Window**

Figure 3.1.1 Screenshot of Main Application Window

*3.3 Encrypt or decrypt data while retrieving from hidden volume:*

*3.1.2 Volume Location*

Volume Location Window allows the user to select the file for which user want to create volume.

For our experiment, we use a laptop PentiumV 2.4 GHz CPU, in which performance data is collected. In the experiments, the laptop encrypts a different file size ranges from 321K byte to 7.139Mega Byte. Several performance metrics are collected:

1- encryption time

2- CPU process time

3- CPU clock cycles and battery power.

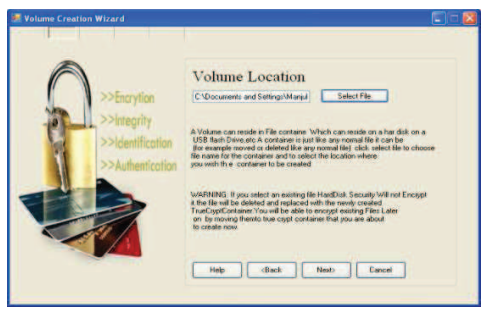

The encryption time is considered the time that an encryption algorithm takes to produce a cipher text from a plaintext. Encryption time is used to calculate the of an encryption scheme. It indicates the speed of encryption. The CPU process time is the time that a CPU is committed only to the particular process of calculations. It reflects the load of the CPU. The more CPU time is used in the encryption process, the higher is the load of the CPU. The CPU clock cycles are a metric, reflecting the energy consumption of the CPU while operating on encryption operations. Each cycle of CPU will consume a small amount of energy.

Figure 3.1.2 Screenshot of Volume Location Window

*3.2  Volume Password*

Volume Password window will allow the user to enter the password and confirm Password.

Password implements user authentication.

4 Simulation Results

The effect of changing key size of AES on power consumption. The performance comparison point is the changing different key sizes for AES algorithm. In case of AES, We consider the three different key sizes possible. In case of AES it can be seen that higher key size leads to clear change in the battery and time consumption. It can be seen that going from 128 bits key to 192 bits causes increase in power and time consumption about 8% and to 256 bit key causes an increase of 16% [9]. The simulation results with different key sizes are as shown in Table2.

| AES Key Size | AES 128 | AES 128 | AES 256 |
|---|---|---|---|
| Time in Milliseconds | 287 | 310 | 330 |

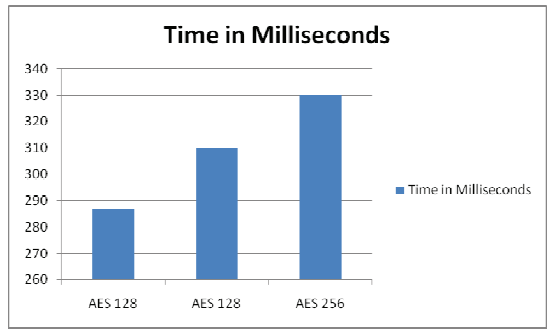

**Figure3.3.1 Time with different key size**

## 6 Conclusion

Full disk encryption (FDE) appears to offer an ideal solution to the losses of data on laptops, CDs and thumb drives. The DiskTrust technique proposed in the paper instead encrypting the entire contents of disk, it encryupts a data stored on single volume which less expensive in terms of time & cost. Disktrust provides authentication, integrity and confidentiality for stored data. DiskTrust implements Symmetric key cryptography with AES. AES is more promising according to results.